\documentclass[twocolumn]{aastex631}

\newcommand\pks{PKS~1222+216}
\newcommand\fermi{{\it Fermi}}
\newcommand\gr{$\gamma$-ray}

\shorttitle{Polarized optical emission of PKS 1222+216}
\shortauthors{Zhang \& Wang}
\graphicspath{{./}{figures/}}

\begin{document}

\title{Polarized Optical Emission of the Blazar PKS 1222+216: Discovery
of A 420-day Quasi-Periodic Signal}

\author{Pengfei Zhang}
\affiliation{Department of Astronomy, School of Physics and Astronomy, Key Laboratory of Astroparticle Physics of Yunnan Province, Yunnan University, Kunming 650091, People's Republic of China; zhangpengfei@ynu.edu.cn; wangzx20@ynu.edu.cn}

\author[0000-0002-0786-7307]{Zhongxiang Wang}
\affiliation{Department of Astronomy, School of Physics and Astronomy, Key Laboratory of Astroparticle Physics of Yunnan Province, Yunnan University, Kunming 650091, People's Republic of China; zhangpengfei@ynu.edu.cn; wangzx20@ynu.edu.cn}
\affiliation{Shanghai Astronomical Observatory, Chinese Academy of Sciences, 80 Nandan Road, Shanghai 200030, China}



\begin{abstract}
We report our search for quasi-periodic signals in long-term optical 
and $\gamma$-ray data for the blazar PKS~1222+216, where the data 
	are from the Steward Observatory blazar monitoring program and 
	the all-sky survey with the Large Area Telescope onboard the {\it
	Fermi Gamma-ray Space Telescope}, respectively. 
	A quasi-periodic signal, with a period of $\simeq$420\,days 
	and a significance of $>5\sigma$, is found in the measurements 
	of the optical linear polarization 
	degree for the source, while no similar signals are found in 
	the optical and $\gamma$-ray light curves 
	covering approximately
	the same time period of $\sim$10\,yr. We study the 
	quasi-periodic variations by applying a helical jet model and find
	that the model can provide a good explanation. This work shows that
	polarimetry can be a powerful tool for revealing the physical
	properties, in particular the configuration of the magnetic fields 
	of jets from galactic supermassive black holes.
\end{abstract}



\section{Introduction}
\label{sec1:intro}

In recent years, the availability of rich data over nearly the whole
electromagnetic spectrum has greatly enabled studies of variable and transient
phenomena in our universe. A particular and interesting one is quasi-periodic 
oscillation (QPO) seen in Active Galactic Nuclei (AGN). The QPO phenomenon 
in the past has
been frequently seen in X-ray observations of Galactic X-ray 
binaries (e.g., \citealt{im19} and references therein).
Now as long-term 
monitoring data at multiple wavelength bands are available, allowing 
searches for different time-scale variability,
quite a few long-periodicity QPOs have been reported
(see, e.g., \citealt{zw21} and references therein). 

Given the periodicties and likely transient nature (i.e., only appearing
during certain time periods), AGN QPOs are discussed to reflect the accretion
activity in the innermost stable circular orbit around an SMBH \citep{gie+08}, 
the quasi-periodic variations of the accretion disk surrounding an SMBH, 
or the binary nature of the SMBH system in the center 
(e.g., \citealt{kin+13,ack+15},
and references therein). As approximately 10\% of AGN have jets and the jets
can contribute with a significant fraction of the observed emission, 
particularly from those so-called blazars (whose jets point close to our line 
of sight), the observed flux variations including QPOs 
can be predominately caused by the variations of the jets. Since the jets are
considered to be coupled with the accretion disk, their variations still 
reflect the physical activity in an AGN.
\begin{figure*}
\centering
\includegraphics[angle=0,scale=0.6]{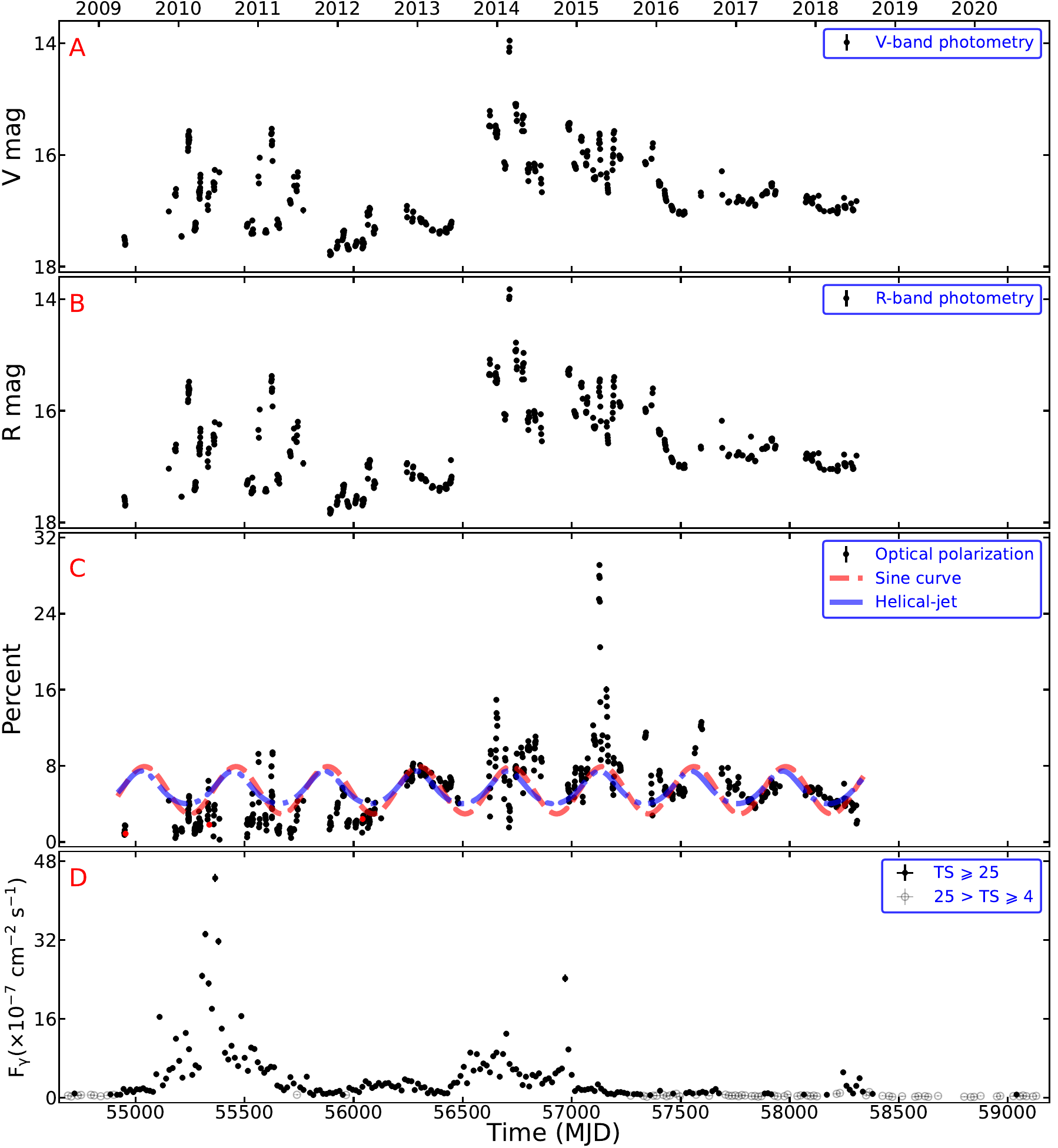}
	\caption{Optical $V$- and $R$-band brightness measurements (panels
	A and B respectively), polarization degree measurements in 
	the wavelength range
	of 5000--7000\,\AA\ (panel C), and 15-day binned \gr\ fluxes in 
	0.1--500\,GeV 
	(panel D) for the blazar \pks\ (data points with 
	TS $\geqslant$ 25 and 25 $>$ TS $\geqslant$ 4 are in black and gray
	respectively).
	The red dashed
	and blue dash-dotted lines respectively indicate a sinusoidal fit 
	and our model fit to the QPO variation (see the text in 
	Sections~\ref{sec:op} and \ref{sec:dis} respectively).
	\label{fig:lc}}
\end{figure*}

The AGN jets are relativistic, with the bulk Lorentz factor 
$\Gamma_b\sim 10$. For blazars, the large and dominant flux variations
are seen arising from their jets, boosted by the Doppler beaming effect.
Thus if a jet has a helical structure, an emitting blob in this jet that
is moving helically would induce periodic flux modulations, as our viewing
angle to the blob is changing periodically (e.g., \citealt{rie04,sss17}).
This geometric model has been applied to a few QPO cases such as
those in the blazar PKS~2247$-$131 \citep{zho+18} and in the narrow-line 
Seyfert~1 Galaxy J0849+5108 \citep{zw21}. Accompanying the flux modulation, 
it is conceivable to think that the polarization of the emission would be 
changing accordingly, as the magnetic field is considered to have a helical
configuration and the synchrotron radiation, which is responsible for emission
in approximately from infrared to X-ray wavelengths from a jet, is highly 
polarized (see e.g.,
discussion in \citealt{cz21}). Indeed, there have been studies and
discussions for the related polarization changes 
(e.g., \citealt{lpg05,rai+13,zha19}, and references therein). 
Previously \citet{ote+20} have even reported a $\gtrsim 4\sigma$ 
QPO signal (with a period of $\sim$1.9\,yr) in both the optical broadband
photometric and polarimetric measurements for the blazar source B2~1633+38.
Thus polarimetry can provide insights into jets' physical properties 
\citep{mjw17}. Moreover if QPOs are detected, they could help revealing the
configuration of the magnetic fields of jets.

In this paper we report another QPO case found
in the polarized optical emission from a blazar, \pks\ (or 4C+21.35).
This source is a flat spectrum radio quasar (FSRQ) subtype
blazar, with redshift $z = 0.434$ \citep{ahn+12}. It has been detected with
the Large Area Telescope (LAT) onboard the {\it Fermi Gamma-ray Space Telescope
(Fermi)} with name J1224.7+2121 in the {\fermi-LAT} first source catalog 
(1FGL; \citealt{1fgl}), and it was selected as a target in
the Steward Observatory blazar monitoring program \citep{smi+09}.
Using the data from the optical spectropolarimetry and photometry carried out
by the program, we conducted the analysis of the source's long-term variations.
The \fermi-LAT 0.1--500\,GeV data for the source were also analyzed, since 
the \gr\ band
might provide additional information for its overall multi-band variations. 
Below we describe the data and analysis in Section~\ref{sec:data},
and present the QPO search results in Section~\ref{sec:res}.
The results are discussed in Section~\ref{sec:dis}.


\begin{figure*}
\centering
\includegraphics[angle=0,scale=0.6]{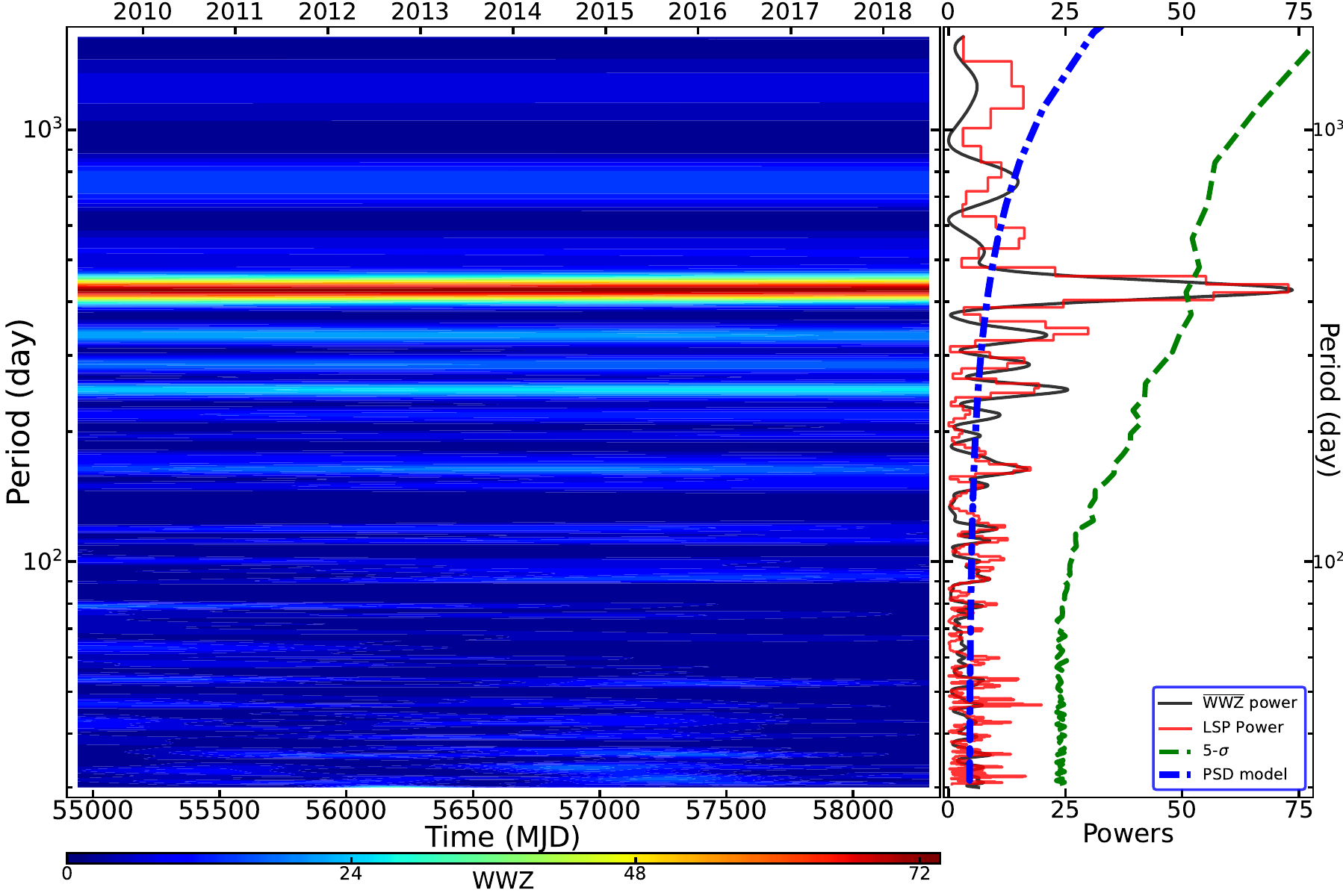}
\caption{Power spectra of the linear polarization degree data in the optical
	wavelength 5000--7000\,\AA\ (cf., panel C in Figure~\ref{fig:lc}),
	showing a periodicity signal at $\sim$420\,day.
	{\it Left} panel: WWZ power over the time range of $\sim$2009--2018.
	{\it Right} panel: time-averaged WWZ power (black line) and LSP power
	(red line). The best-fit to the underlying LSP power spectral density
	is shown as the blue dash-dotted line. The light curve simulation 
	yields a 5$\sigma$ significance curve, shown as the green dashed line.
}
\label{fig:powp}
\end{figure*}

\section{Data and Analysis }
\label{sec:data}

\subsection{Optical data description}

Nearly 100 blazars were monitored by the Steward Observatory for 10 
\emph{Fermi} mission cycles from 2008 October to 2018 July. \pks\ was
one of the targets and for it the linear polarization measurements at 
a wavelength range of 5000--7000\,\AA\ and photometric
$V$- and $R$-band 
brightness measurements are available.
The time range of the data extends approximately between MJD~54948--58306.

We used the degree of polarization (DoP) measurements, whose minimum 
(at MJD 55385.18), maximum (at MJD 57127.25), and mean values in 
percentage are 0.25, 29.1, and 5.47, respectively, and standard deviation 
is 366.45\%.
The photometric $V$- and $R$-band brightness measurements are simultaneous, 
and the $V$-band ($R$-band) magnitudes have a minimum
(at MJD 56715.39), maximum (at MJD 55892.51),
and mean value of 13.95 (13.82), 17.79 (17.84), and 16.64 (16.58),
respectively. The standard deviation is 69.69\% (76.28\%) in $V$-band 
($R$-band).
The DoP and magnitude data are shown in Figure~\ref{fig:lc}, and their median 
and mean time intervals between two adjacent data points 
are $\sim$1 and 7 days, respectively. It can be noted that there is an 
approximate three-month gap in the data of each year, obviously because of
the observations being ground-based.

\subsection{\emph{Fermi}-LAT data analysis}

In the \fermi-LAT fourth source catalog (4FGL; \citealt{abd+20,dr3}), 
PKS 1222+216 is named 4FGL J1224.9+2122 with coordinate 
R.~A.~=~$\rm12^h24^m54^s.46$ and decl.~=~$21^{\circ}22'46''.39$
We selected the \emph{Fermi}-LAT 0.1-500 GeV Pass 8 \emph{Front}+\emph{Back} 
SOURCE class photon-like events between 2008 August 4 and 2021 November 18 
in a region of 20$^\circ\times$20$^\circ$ (region of interest; RoI), 
centered at the position of PKS 1222+216.
We removed the events with zenith angle $>90^{\circ}$, and reduced the others 
with expression DATA\_QUAL$>$0~\&\&~LAT\_CONFIG~=~1 to obtain high-quality 
data in the good time intervals.

A model file was created with the script make4FGLxml.py\footnote{https://fermi.gsfc.nasa.gov/ssc/data/analysis/user/} based on 4FGL,
which included best-fit spectral parameters of all known 4FGL sources 
in the RoI.
We set a few parameters free, which were the parameters of flux 
normalizations and spectral shapes for the sources within 5$^\circ$ from 
the target, and the normalizations for the sources within 
5$^\circ$--10$^\circ$ and the ones outside 10$^\circ$ but identified with 
significant variations (i.e., their \emph{Variablility\_Index}~$\geqslant72.44$).
The normalizations of the Galactic and extragalactic diffuse emission components
were also set free.
The other parameters were fixed at the values given in 4FGL.

A binned maximum likelihood analysis was 
first employed to build a best-fit model file for the RoI.
PKS~1222+216 has a log-parabola spectral shape in 4FGL,
$dN/dE\sim (E/E_b)^{-\alpha_{s}-\beta_{s}\log(E/E_b)}$. From the likelihood
analysis, we obtained a test statistic (TS) value of $\sim$83,470 and
an average photon flux of 
$(267.85\pm2.12)\times10^{-9}$~photons~cm$^{-2}$~s$^{-1}$ in 0.1$-$500\,GeV 
band from the whole data.
The best-fit spectral parameters were $\alpha_s=2.324\pm0.007$, 
$\beta_s=0.029\pm0.004$, and $E_b=392.380\pm10.219$\,MeV. The values
are in well agreement with those reported in the 4FGL data release 3 (DR3; 
\citealt{dr3}). Based on the best-fit model file, a 0.1--500\,GeV light curve 
of PKS 1222+216 was constructed by performing an unbinned likelihood analysis.
In this step, we fixed the parameters of spectral shapes of the 
sources at their best-fit values obtained above and set the normalizations
of each sources in the RoI free.
We tested different time bins, such as 15, 30, and 45\,day, but the obtained
light curves did not have significant differences.
In the bottom panel of Figure~\ref{fig:lc}, the 15-day binned light curve 
is presented. In order to show the light curve as complete as possible, 
low-significant data points with 25 $>$ TS $\geqslant$ 4 are also included.


\section{Periodic variability analysis and results}
\label{sec:res}

We searched for periodicity signals in the optical and \gr\ data.
Two methods were used, namely the weighted wavelet Z-transform\footnote{https://github.com/eaydin/WWZ}
\citep[WWZ;][]{fos96} and
the generalized Lomb-Scargle Periodogram\footnote{https://docs.astropy.org/}
 \citep[LSP;][]{lom76,sca82,zk09}. The
latter was used to check the results from the former
as an independent method.

\subsection{Optical polarization}
\label{sec:op}

In the optical DoP data, we found a $\sim$420-day periodic signal.
As clearly seen in the WWZ time-frequency power spectrum shown in the
left panel of Figure~\ref{fig:powp},
the signal was present over the whole-data time range and is significant in
the time-averaged power spectrum (right panel of Figure~\ref{fig:powp}).
The LSP power spectrum shows a similar result, containing a sharp 
peak at the same period.
Here we adjusted the parameter for the wavelet transform in the
WWZ analysis to make the two power spectra as much as possible match 
with each other.
We estimated the period value by fitting the power peak
with a Gaussian function, which yielded 420.1$\pm$55.6\,day, where 
the uncertainty was taken as the full width at half maximum of the peak.

A periodic signal or statistical fluctuations in a set of time series data 
may lead to a power peak in an LSP power spectrum.
The probability ($p$) for the power peak arising purely from 
the statistical fluctuations was estimated to be 
$<7.8\times10^{-35}$ (with respect to the local noise floor around 
the peak),
according to the calculation method given in \citet{hb86} and \citet{zk09}.
The false alarm probability (FAP) considering the period search in 
the frequency range is $<8.4\times10^{-33}$, given by 
$\mathrm{FAP}=1-(1-p)^{N}$, where $N$ (=108) is the number of independent 
frequencies in the frequency range (i.e., the trial factor) and 
calculated from
$(f_{\rm max} - f_{\rm min})/\delta_f$. Since the time series data are
unevenly spaced, we approximately set 
$f_{\rm max} \simeq 1/(30\,{\rm day})$ and
$f_{\rm min}\simeq 1/(1640\,{\rm day})$ (where 1640\,day $\sim$ 
half of the total length of the data), 
and $\delta f$ is the frequency resolution, determined by
the total length of the data.

\begin{figure*}[!ht]
\centering
\includegraphics[angle=0,scale=0.45]{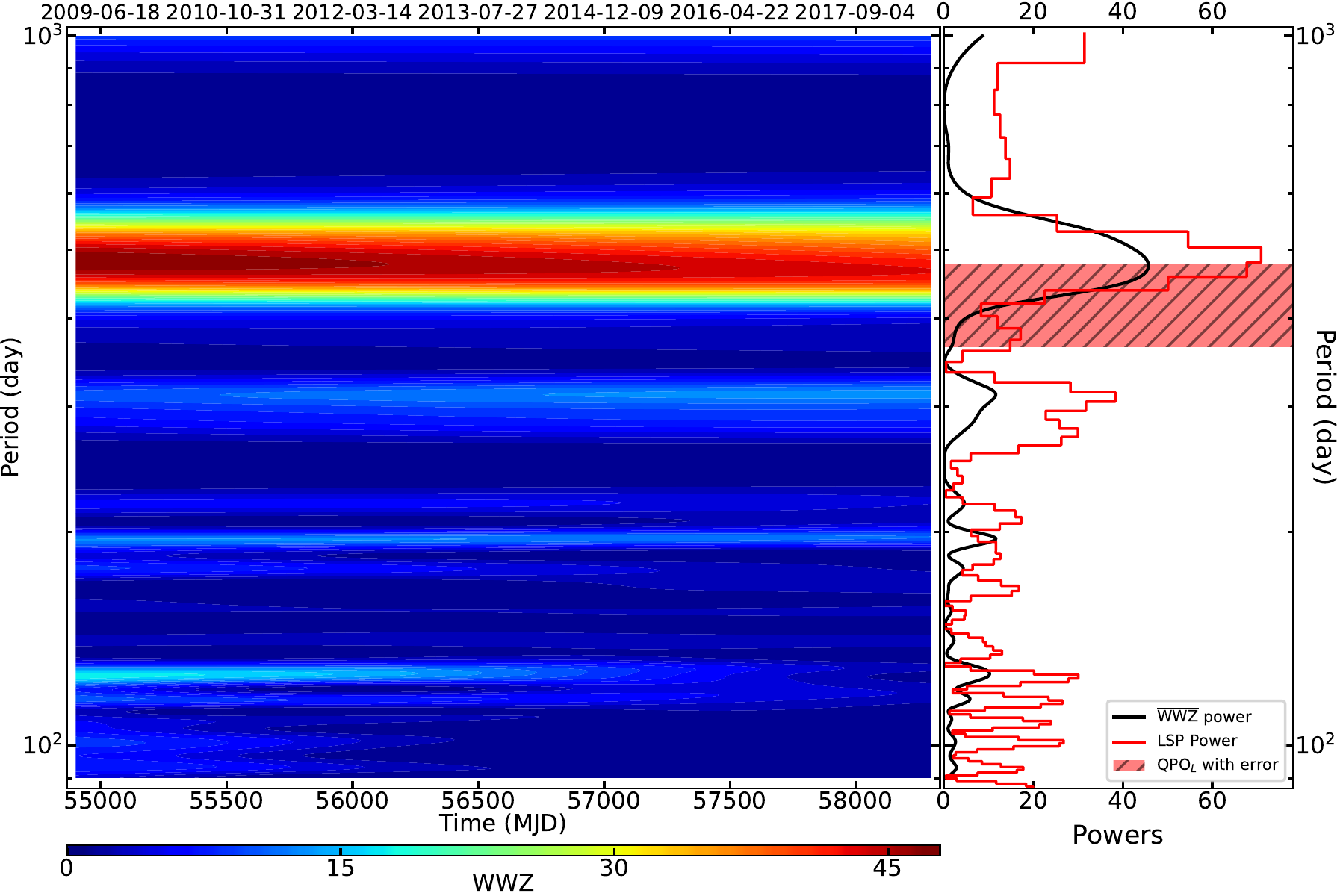}
\includegraphics[angle=0,scale=0.45]{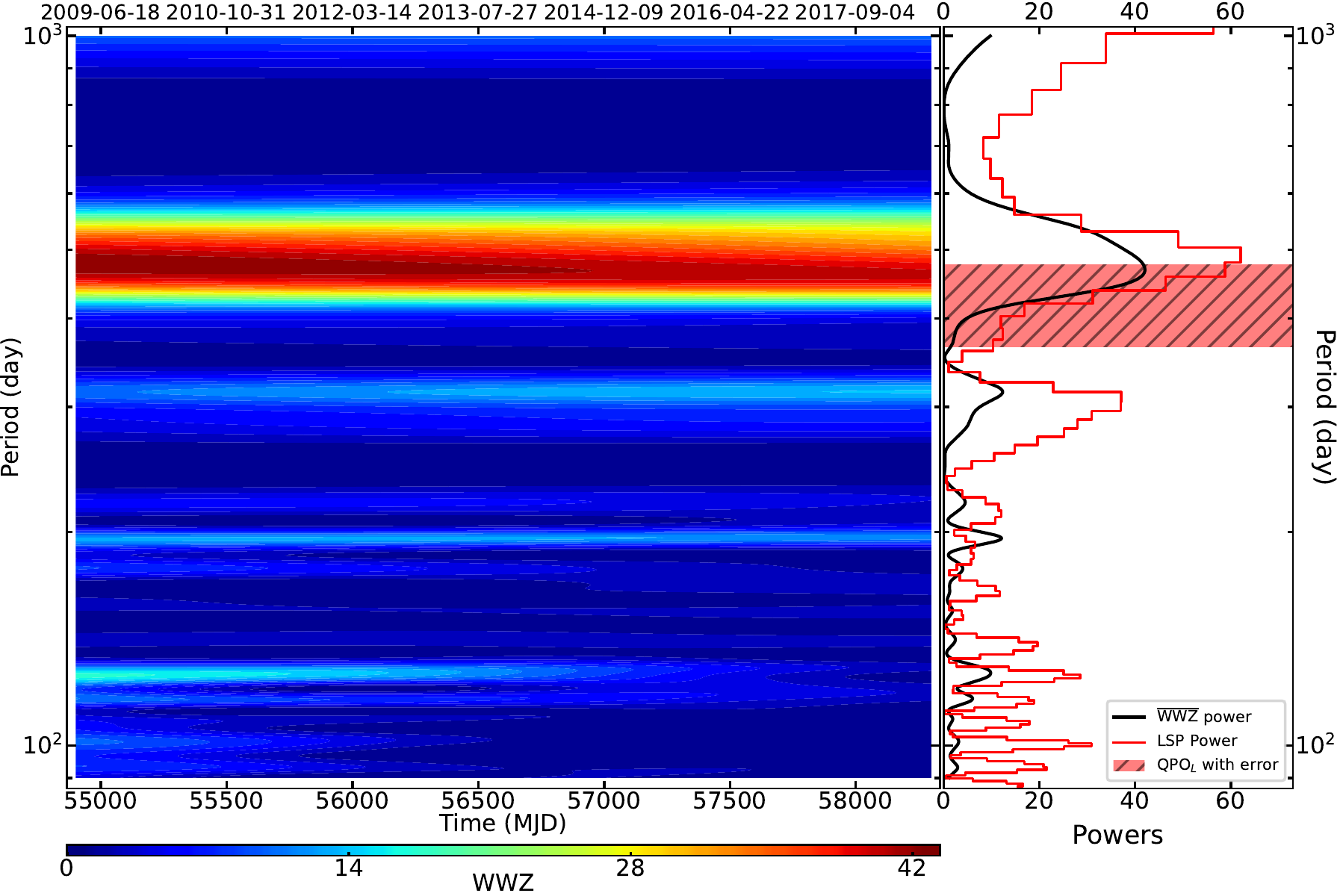}
\includegraphics[angle=0,scale=0.45]{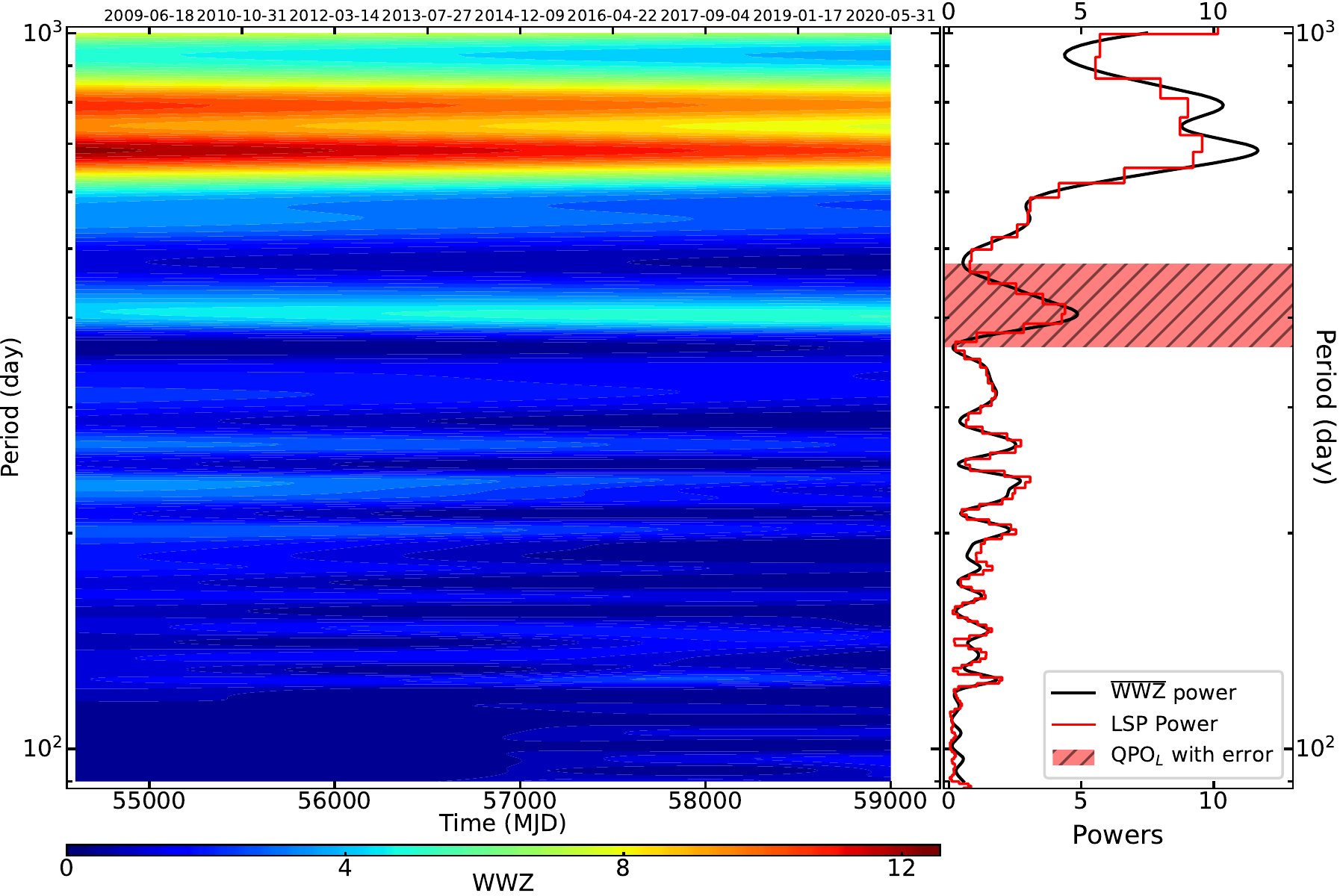}
\caption{Power spectra from the light curves of \pks\ at optical 
	$V$- and $R$-band 
	from the Steward Observatory observations ({\it top left} and 
	{\it top right} respectively) and at $\gamma$-rays from 
	the \emph{Fermi}-LAT
	observations ({\it bottom}). The red-black shaded region in each panels
	marks the uncertainty range of the 420-day periodicity found in 
	the optical polarization data. Other marks are the same as those in 
	Figure~\ref{fig:powp}.}
\label{fig:powo}
\end{figure*}

We employed a light-curve simulation method to determine the significance 
of the periodicity signal.  The LSP power spectrum was modeled with a function 
of a smoothly bending power law plus a constant \citep{gv12}
to estimate the underlying power spectral density (PSD). The function has 
the form of $P(f)=K f^{-\alpha}\left[1+\left(f / f_{\text {bend }}\right)^{\beta-\alpha}\right]^{-1}+C$,
where $K, \alpha, \beta, f_{bend}$, and $C$ are the normalization, 
low frequency slope below bending, high frequency slope above bending, 
bending frequency, and Poisson noise, respectively.
Their best-fit parameters were obtained with a maximum likelihood algorithm 
provided in \citet{bv12}, and we found
$K=0.56\pm0.80$, $\alpha=0.66\pm0.28$, $\beta=1.56\pm0.73$, 
$f_{bend}=0.0003\pm0.0023$, and $C=4.43\pm0.69$ (the model fit is shown
in the right panel of Figure~\ref{fig:powp}).
Based on this model, $2\times10^7$ artificial light curves were 
generated \citep{emp13}.
For each artificial light curve, we obtained its PSD with the above LSP
periodic analysis.
Significance curves were calculated by counting the PSD data points at each of
the frequencies.
The resulting 5$\sigma$ significance curve is shown in 
the right panel of Figure~\ref{fig:powp}, and as can be seen,
the LSP power peak of the 420-day periodicity exceeds 
this significance curve.
After using the same method as the above for 
taking into account the trial factor,
the signal is still greater than 5$\sigma$.
As to a few minor peaks also seen in the power
spectra, we found that none of them had a sufficiently high significance, 
particularly when the trial factor was taken into account.

We considered a sinusoidal modulation curve for approximately describing 
the periodic variations,
$A \sin \left[2 \pi f\left(t-t_{0}\right)\right]+A_{0}$,
where $A, f, t_0,$ and $A_0$ are the amplitude, frequency, 
time of zero phase, and additive offset of the sine function, respectively.
The values of the best-fit parameters were provided by the LSP code,
which were $2.49\pm0.18$, 
$0.00238\pm0.00002$ day$^{-1}$, $54935.54\pm4.81$, and $5.44\pm0.13$, 
respectively.
This sinusoidal curve is shown in the panel C of Figure~\ref{fig:lc}.
Approximately 8 cycles of periodic modulation appeared in the data.

Although the 420-day value is different from a year of 365\,day, we 
investigated whether
the yearly observational gaps could induce any artifactual signals.
We conducted two simulations by
generating DoP data points with their times set as the observational ones. The
DoP values were set as a constant (5.47, the mean of the observed data),
or random ones in a range of 0--15 (the DoP values of most of the
data points).
In the first simulation, there were small peaks in the power spectrum, 
but none
of them appears significant and more importantly around 420\,day. 
In the second
simulation, there were only background noises. We thus concluded that
the 420-day signal could not be induced by the yearly observational gaps.


\subsection{Optical and \gr\ light curves}

The same analysis for periodicity search was conducted to 
the optical photometric $V$- and $R$-band data.
The obtained power spectra are shown in Figure~\ref{fig:powo}. In the power 
spectra resulting from both bands,
a possible period of $\sim$500-day appears.
While it can be considered to overlap with 
that from the DoP data, it is broad and not significantly among other nearby
weak peaks.

The analysis on the \fermi-LAT \gr\ data yielded similar results to the above
optical data (Figure~\ref{fig:powo}). A weak signal appears, with its
periodicity consistent with the error range found for the 420-day periodicity,
but compared with nearby peaks, it is not significant at all.
We concluded that no significant signals were found in either the optical
or \gr\ light curve data.


\section{Discussion}
\label{sec:dis}

Taking advantage of the availability of long-term monitoring data at different
wavelength bands, we have analyzed the optical and \gr\ data taken for 
the blazar \pks. In the 10-year long optical photometric
and polarimetric data, we have found a QPO signal of period 
$P_{\rm QPO}\simeq$420\,day in the DoP measurements of the latter. 
While there are peaks at $\sim P_{\rm QPO}$ 
present in the power spectra derived from
the optical photometric and \fermi-LAT \gr\ data, they are too weak to
be considered as possible signals. Also very-high-energy emission from
the source was detected with the MAGIC 
telescopes in the energy range of 70~GeV up to 
$\sim$400~GeV
during its active high-energy \gr\ variation period \citep{ale+11}, 
and multi-wavelength 
studies of the source have
clearly indicated that the major \gr\ variations, such as that around
MJD~55300 (Figure~\ref{fig:lc}), were flaring 
events \citep{ack+14,bha+21}. Non-detection of periodic signals 
should be expected when the \gr\ flux variations were dominated 
by flaring activity.

QPO activity seen from a blazar should arise from its jets, no matter 
what the underlying cause is, for example due to a binary SMBH system or 
a precessing accretion disk. 
This QPO in the optical polarized emission of a blazar presents 
an interesting case
among thus-far reported QPOs, as nearly all of them were found in flux 
variations. Different from most of the others, this QPO possibly reveals the  
properties of the magnetic field in the source's jet \citep{rai+13,mjw17}.
A geometric model has been applied to explain the often-reported quasi-periodic flux variations considering a helical jet and magnetic field \citep{cz21}.
An emitting blob in such a jet moves along the helical path, causing periodic
changes in our view angle $\theta$ to the jet and as a result in
the observed flux, since the Doppler beaming factor is a function of $\theta$
(e.g., \citealt{zho+18}).

Similar scenarios have been proposed to explain the polarization variations
seen in the synchrotron emission of blazars \citep{lpg05}, where helical 
magnetic fields are considered. Following \citet{rai+13}, we assume that
DoP $P$ changes as a function of the viewing angle $\theta'$
in the jet rest frame, $P=P_{\rm max}\sin^2\theta'$, where $P_{\rm max}$ is the 
maximum DoP value and is shown to be $\sim 20$ in \citet{lpg05}.
We tested to fit the DoP measurements with this function,
for which $\sin\theta'=\sin\theta/\Gamma_b(1-\beta\cos\theta)$ ($\beta c$ is 
the bulk speed). The viewing angle $\theta$ is derived from
$\cos\theta = \sin\phi\sin\psi\cos(2\pi t/P_{\rm QPO})+\cos\phi\cos\psi$, where
$\phi$ is the pitch angle of an emitting blob's motion from the jet and
$\psi$ is the inclination angle of the jet to the line of sight. According to
the detailed studies of the components in the jet at the center of \pks\ with
the Very Long Baseline Array (VLBA) observations \citep{jor+17}, the opening
semi-angle of its jet is $\sim 2\arcdeg$ and thus we set $\phi=2\arcdeg$.
In our fitting, we also fixed $\Gamma_b=15$, which was obtained from the
radio measurements and the spectral energy distribution modelling reported in
\citealt{jor+17} and \citealt{bha+21} respectively. Since polarized
emission of a jet can consist of multiple components 
(e.g., \citealt{mjw17}), 
we did not intend to fit the observed DoP values exactly and thus
set $P_{\rm max}$ as a free parameter. The other one set free was 
$\psi$. 
We obtained the best fit with
$P_{\rm max}=9.37\pm0.03$ and $\psi=8.22\pm0.02$\,deg.
This fit,
shown in Figure~\ref{fig:lc}, can describe the overall DoP variations and is
similar to the sinusoidal fit given by the LSP code.
In this geometric model, the view angle changes
around $\psi\simeq 8\arcdeg$, which is close to the average value $5.6\arcdeg\pm1.0\arcdeg$ estimated from the VLBA studies \citep{jor+17}. 
Therefore the model provides
a reasonable explanation to the QPO seen in the DoP measurements, in which
the magnetic field of the jet should have a helical configuration.

We note that the nearly simultaneous optical light curves,
unlike what was reported for the B2~1633+38 case \citep{ote+20},
do not show
a QPO signal or similar variation patterns (cf., Figure~\ref{fig:lc}).
This can be explained as that the optical emission contains a significant
contribution arising from the accretion disk, as the fitting results from the
studies of the broadband spectral energy distribution have shown that 
the disk emission could be dominant \citep{ack+14,bha+21}. Alternatively,
if the optical emission mainly arises from the jet, the magnetic field could
consist of a turbulent component,
which then explains the observed DoP values $\sim$5\%, much lower
than that expected from the synchrotron
emission of highly ordered magnetic fields \citep{mjw17}.
This case of the QPO signal thus likely reveals a significant helical 
component of the magnetic field in the jet of \pks, and 
shows that polarimetry can be a powerful tool for studies of blazar jets
by overcoming apparently noisy brightness variations.

According to the geometric model, the jet of \pks\ was moving towards 
the Earth along a helical path with a cycle time of 
$[P_{\rm QPO}/(1+z)]/(1-\beta\cos\phi\cos\psi)\simeq 62$\,yr at the 
host galaxy (cf., \citealt{zho+18}). We observed $\sim$8 QPO cycles in 10\,yr, 
which
implies that the emitting blob moved with a distance of $\sim$150\,pc in total
or a projected distance of $\sim$21\,pc. The luminosity distance to \pks\
is estimated to be 2480\,Mpc (where the cosmological parameters
from the Planck mission, $H_0\simeq 67.4$\,km\,s$^{-1}$\,Mpc$^{-1}$, were used;
\citealt{pla2020}).
Thus the corresponding projected angular distance would be 
$\sim$1.8\,mas.
Comparing to the proper motion values of 
0.01--0.5\,mas\,yr$^{-1}$ derived from the VLBA observations for several 
components of the jet in
\pks\ \citep{jor+17}, the estimated projected distance is within the range, 
showing the consistency of the model result with that of the high-resolution 
radio imaging.

As a summary, we have found a $\simeq$420-day periodicity in the optical
linear DoP measurements for the blazar \pks. The QPO signal is
significant at a $>5\sigma$ level, higher than that previously reported in
B2~1633+38. We have applied a helical jet model to explain the quasi-periodic 
variations, and the parameters estimated from the model are consistent with
those derived from the VLBA studies of the central radio jet.
This QPO case
shows that while the flux variations of a blazar may appear random, the
physical properties of its jets can still be investigated and revealed through
polarimetry. It is challenging to carry out polarimetry observations because
of the need for detecting sufficient amount of photons, but when
QPOs shown in flux variations of blazars are found, close polarimetry 
monitoring should be warranted as the observation can potentially provide 
critical information for understanding the mechanism that drives the QPOs 
and thus jets' physical properties.

\begin{acknowledgments}

Data from the Steward Observatory blazar monitoring program were 
used. This program was supported by Fermi Guest Investigator grants NNX08AW56G, 
NNX09AU10G, NNX12AO93G, and NNX15AU81G.

We are very grateful for the referee's detailed and helpful comments.
This work is supported in part by the National Key R\&D Program of China 
under grant No. 2018YFA0404204, the National Natural Science Foundation of 
China No.~12163006, the Basic Research Program of Yunnan Province 
No. 202101AT070394, and the joint foundation of Department of Science and 
Technology of Yunnan Province and Yunnan University [2018FY001 (-003)].
Z.W. acknowledges the support by the Original
Innovation Program of the Chinese Academy of Sciences (E085021002) and
the Basic Research Program of Yunnan Province No. 202201AS070005.

\end{acknowledgments}




\bibliographystyle{aasjournal}



\end{document}